\documentstyle[twocolumn,aps,amssymb,epsfig]{revtex}
\begin{document}


\def\calA{{\cal A}}
\def\calH{{\cal H}}
\def\calL{{\cal L}}
\def\calO{{\cal O}}

\def\vecp{{\vec p}}

\def\etal{{\it et al.}}
\def\ibid#1#2#3{{\it ibid}. {\bf #1}, #3 (#2)}

\def\epjc#1#2#3{Eur. Phys. J. C {\bf #1}, #3 (#2)}
\def\ijmpa#1#2#3{Int. J. Mod. Phys. A {\bf #1}, #3 (#2)}
\def\jhep#1#2#3{J. High Energy Phys. {\bf #1}, #3 (#2)}
\def\mpl#1#2#3{Mod. Phys. Lett. A {\bf #1}, #3 (#2)}
\def\npb#1#2#3{Nucl. Phys. {\bf B#1}, #3 (#2)}
\def\plb#1#2#3{Phys. Lett. B {\bf #1}, #3 (#2)}
\def\prd#1#2#3{Phys. Rev. D {\bf #1}, #3 (#2)}
\def\prl#1#2#3{Phys. Rev. Lett. {\bf #1}, #3 (#2)}
\def\rep#1#2#3{Phys. Rep. {\bf #1}, #3 (#2)}
\def\zpc#1#2#3{Z. Phys. {\bf #1}, #3 (#2)}

\twocolumn[\hsize\textwidth\columnwidth\hsize\csname
@twocolumnfalse\endcsname

\title{Photon polarization with anomalous right-handed top couplings in 
$B\to K_{\rm res}\gamma$}
\author{Jong-Phil Lee}
\address{Department of Physics and IPAP, Yonsei University, Seoul, 120-749, Korea\\{\rm e-mail: jplee@phya.yonsei.ac.kr}}

\tighten
\maketitle

\begin{abstract}
The effect of anomalous right-handed top couplings on the photon polarization in
$B\to K_{\rm res}\gamma$ is investigated.
It is recently reported that the photon polarization can be measured through 
the up-down asymmetry of the photon direction relative to the subsequent 
$K_{\rm res}$ decay plane.
We find that the anomalous couplings can severely affect the photon 
polarization without spoiling the well measured branching ratio of 
$B\to X_s\gamma$.
Different features from other scenarios are also discussed.
\end{abstract}
\pacs{}
\pagebreak
]


Radiative $B$ decay of $B\to X_s\gamma$ plays important rolls in testing the
standard model (SM) and constraining the new physics.
Not only the inclusive and exclusive branching ratios but also polarization of
the emitted photon have been extensively studied
\cite{Song,Atwood,Mannel,Melikhov}.
In the SM, photons from $b\to s\gamma$ are predominantly left handed up to 
$\calO(m_s/m_b)$.
A smart way of measuring the photon polarization in $B\to K_{\rm res}\gamma$ is
recently proposed \cite{Gronau}.
In this approach, hadronic three body decay of 
$K_{\rm res}\to K^*\pi~({\rm or}~\rho K)\to K\pi\pi$ is essential to construct
a triple vector product $\vecp_\gamma\cdot(\vecp_1\times\vecp_2)$.
Here $\vecp_\gamma$ is the photon momentum and $\vecp_1$, $\vecp_2$ are two of
the daughter hadron momenta, measured in the $K_{\rm res}$ rest frame.
A remarkable result is that the so called photon polarization parameter 
$\lambda_\gamma$, which encodes the left-right asymmetry of the emitted photon
polarization, is universally determined by the Wilson coefficients of the 
effective Hamiltonian.
In the SM, the predominant left-handedness implies that 
$\lambda_\gamma= -1+\calO(m_s^2/m_b^2)$.
The authors of \cite{Gronau} predicted that the integrated up-down asymmetry
in $K_1(1400)\to K\pi\pi$ is $(0.33\pm0.05)\lambda_\gamma$, which is quite
large compared to other $K_{\rm res}$ decays.
Estimated number of $B{\bar B}$ pairs required to measure the asymmetry is 
about $10^8$, which is already within the reach of current $B$ factories.
A stringent test of SM as well as constraints on new physics by the photon
polarization is therefore near at hand.
\par
In this Letter, we investigate the effects of anomalous right-handed couplings
on the photon polarization parameter.
The left-right (LR) symmetric model and the minimal supersymmetric standard 
model (MSSM) can provide such new couplings.
The LR model is one of the natural extension of the SM, based on the 
$SU(2)_L\times SU(2)_R\times U(1)$ gauge group.
Besides the usual left-handed quark mixing, right-handed quark mixing is also
possible in the LR model.
Without a manifest symmetry between the left- and right-handed sectors, the
right-handed quark mixing is not necessarily the same as the left-handed quark
mixing governed by the CKM paradigm.
Consequently, there are additional right-handed charged current interactions 
with couplings different from the left ones, which are suppressed by the heavy
extra $W$ boson \cite{Babu}.
\par
In the unconstrained MSSM (uMSSM), the gluino-involved loop can contribute to 
the "wrong" chirality operator through the left-right squark mixing 
\cite{Everett}.
There can be a special case where $W$, Higgs, chargino, and gluino 
contributions to the ordinary Wilson coefficient $C_{12}$ tend to cancel each
other while the "wrong" chirality coefficient $C_{12}'$ gives the dominant
contribution.
\par
In the present analysis, however, we just concentrate on the anomalous 
right-handed top quark couplings ${\bar t}bW$ and ${\bar t}sW$, ignoring
the effects of additional left-handed interactions and new particles, and not
specifying the underlying models.
The anomalous top couplings deserve much attention because the Large Hadron
Collider (LHC) at near future will produce about $10^7$-$10^8$ top quark pairs
per year.
The ${\bar t}bW$ vertex will be directly measured with high precision through
the dominant $t\to bW$ channel.
On the other hand, the ${\bar t}sW$ vertex is quite suppressed because of the 
small CKM element $|V_{ts}|$.
While the branching ratio is estimated to be 
${\rm Br}(t\to s W)\sim 1.6\times 10^{-3}$,
the large number of top quarks at LHC will enable us also to measure the 
$t\to sW$ process and provide a chance to probe the ${\bar t}sW$ coupling 
directly.
\par
We introduce dimensionless parameters $\xi_s$ and $\xi_b$ where the anomalous
right-handed ${\bar t}sW$ and ${\bar t}bW$ couplings are encapsulated, 
respectively.
Up to the leading order of $\xi$, ordinary Wilson coefficients are modified
through the loop-function corrections proportional to $\xi_b$.
On the other hand, there appear new chiral-flipped operators in the effective 
Hamiltonian.
The corresponding new Wilson coefficients are proportional to $\xi_s$, with
new loop functions.
It was recently shown that there is a parameter space of $(\xi_b, \xi_s)$ 
where the discrepancy of $\sin 2\beta$ between $B\to J/\psi K$ and $B\to\phi K$
is well explained while satisfying the $B\to X_s\gamma$ constraints \cite{KYL}.
Though the allowed values of $\xi$ are rather small, 
modified Wilson coefficients
of the colormagnetic and electromagnetic operators $O_{11}$, $O_{12}$ involve 
large enhancement factor of $m_t/m_b$.
Fortunately, the photon polarization parameter $\lambda_\gamma$ depends only on
the Wilson coefficient $C_{12}$ and its chiral-flipped partner $C'_{12}$, 
irrespective of the species of $K_{\rm res}$.
Thus the photon polarization parameter $\lambda_\gamma$, or the up-down
asymmetry of the emitted photon is quite sensitive to the new right-handed
couplings.
\par
Let us first consider the effective Lagrangian containing possible right-handed
couplings
\begin{equation}
\calL=-\frac{g}{\sqrt{2}}\sum_{q=s,b}V_{tq}
 {\bar t}\gamma^\mu(P_L+\xi_q P_R)q W^+_\mu+{\rm h.c.}~,
\end{equation}
where $P_{L,R}$ are the usual chiral projection operators.
With new dimensionless parameters $\xi_{b,s}$, the effective Hamiltonian for
the radiative $B$ decays has the form of
\begin{eqnarray}
\calH_{\rm rad}&=&-\frac{4G_F}{\sqrt{2}}V_{ts}^*V_{tb}\Big[
 C_{12}(\mu)O_{12}(\mu)+C'_{12}(\mu)O'_{12}(\mu)\Big]~,\nonumber\\
O_{12}&=&\frac{e}{16\pi^2}m_b{\bar s}P_R\sigma_{\mu\nu}bF^{\mu\nu}~,
\end{eqnarray}
and $O_{12}'$ is the chiral conjugate of $O_{12}$.
After matching at $\mu=m_W$, the Wilson coefficients are given by, in the SM,
\cite{Inami}
\begin{eqnarray}
C_{12}(m_W)&=&F(x_t)\nonumber\\
 &=&\frac{x_t(7-5x_t-8x_t^2)}{24(x_t-1)^3}
 -\frac{x_t^2(2-3x_t)}{4(x_t-1)^4}\ln x_t ~,\nonumber\\
C_{12}'(m_W)&=&0~,
\end{eqnarray}
where $x_t=m_t^2/m_W^2$.
Turning on the right-handed ${\bar t}bW$ and ${\bar t}sW$ couplings, the
Wilson coefficients are modified as 
\begin{eqnarray}
C_{12}(m_W)&\to&F(x_t)+\xi_b\frac{m_t}{m_b}F_R(x_t)~,\nonumber\\
C_{12}'(m_W)&\to&\xi_s\frac{m_t}{m_b}F_R(x_t)~,
\end{eqnarray}
with the new loop function \cite{KYL,Cho}
\begin{equation}
F_R(x)=\frac{-20+31x-5x^2}{12(x-1)^2}+\frac{x(2-3x)}{2(x-1)^3}\ln x~.
\end{equation}
Scaling down to $\mu=m_b$ is accomplished by the usual renormalization group
(RG) evolution.
We use the RG improved Wilson coefficients of \cite{KYL}.
\par
Next consider the radiative $B$ decay of 
${\bar B}\to{\bar K}_{\rm res}^{(i)}\gamma$.
The photon polarization in this process is naturally defined as follows:
\begin{equation}
\lambda_\gamma^{(i)}=
\frac{|A_R^{(i)}|^2-|A_L^{(i)}|^2}{|A_R^{(i)}|^2+|A_L^{(i)}|^2}~,
\end{equation}
where 
$A^{(i)}_{L(R)}\equiv\calA({\bar B}\to{\bar K}_{\rm res}^{(i)}\gamma_{L(R)})$
is the weak amplitude for left(right)-polarized photon.
The authors of \cite{Gronau} simply argued that 
\begin{equation}
\langle K_{\rm res}^{(i)R}\gamma_R|O_{12}'|{\bar B}\rangle=(-1)^{J_i-1}P_i
\langle K_{\rm res}^{(i)L}\gamma_L|O_{12}|{\bar B}\rangle~,
\label{LR}
\end{equation}
where $J_i (P_i)$ is the resonance spin (parity).
It means that $|A_R^{(i)}|/|A_L^{(i)}|=|C_{12}'|/|C_{12}|$, and further
\begin{equation}
\lambda_\gamma^{(i)}=\frac{|C_{12}'|^2-|C_{12}|^2}{|C_{12}'|^2+|C_{12}|^2}
\equiv\lambda_\gamma~.
\label{lambdagamma}
\end{equation}
Thus the photon polarization parameter $\lambda_\gamma$ is independent of the
$K_{\rm res}$ states and is universally determined by the Wilson coefficients.
It is also free from the hadronic uncertainty which usually originates from the
weak form factors.
The relation (\ref{LR}) ensures that the common form factors are involved in
$\calA_L$ and $\calA_R$, being canceled in the ratio.
In the SM, $\lambda_\gamma\approx -1$ ($+1$ for ${\bar b}\to{\bar s}\gamma$) 
since $|C_{12}'|/|C_{12}|\approx m_s/m_b$.
\par
To relate $\lambda_\gamma$ with the physical observable, a full analysis
including the strong decays of $K_{\rm res}$ must be implemented.
As mentioned before, there needs at least three hadrons in the final state
to see the up-down asymmetry.
One can readily calculate the angular distribution of 
${\bar B}\to{\bar K}\pi\pi$ \cite{Gronau}, and find that terms containing
the photon polarization parameter $\lambda_\gamma$ are proportional to the
up-down asymmetry of the emitted photon momentum with respect to the
$K\pi\pi$ decay plane.
\par
\begin{figure}
\begin{center}
\epsfig{file=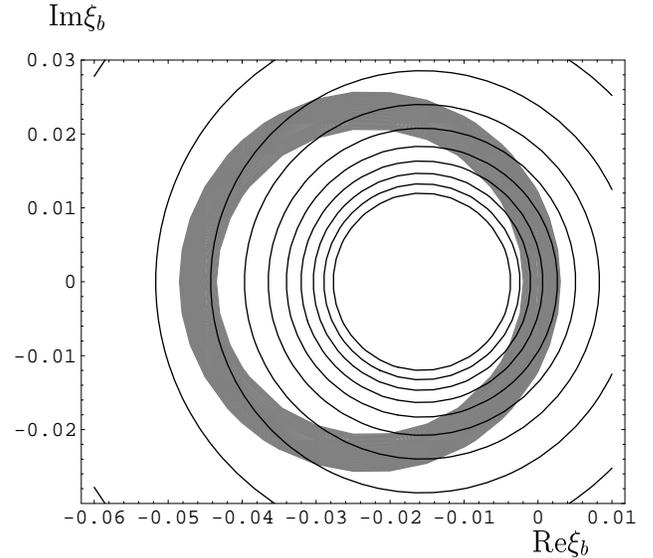,height=8cm}\\
\epsfig{file=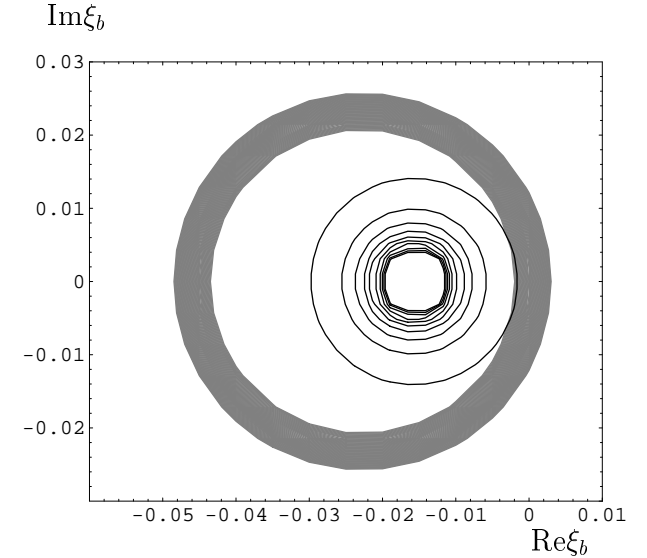,height=8cm}
\end{center}
\caption{Contour plot of $\lambda_\gamma$ for (a) $|\xi_s|=0.012$ and (b)
$|\xi_s|=0.001$. Concentric circles correspond to, from outside to inside, 
$\lambda_\gamma=-0.9$, $-0.8$, $\cdots$, $0$ in (a) and 
$\lambda_\gamma=-0.99$, $-0.98$, $\cdots$ ,$-0.9$ in (b), 
respectively.
Shaded ring is the allowed region from $B\to X_s\gamma$.}
\end{figure}
Now let us examine the effects of newly introduced $\xi_{b,s}$ on 
$\lambda_\gamma$.
New couplings $\xi_{b,s}$ are strongly constrained by the branching ratio
${\rm Br}(B\to X_s\gamma)$ and the $CP$ asymmetry $A_{CP}(B\to X_s\gamma)$.
We use the weighted average of the branching ratio \cite{Everett,KYL}
\begin{equation}
{\rm Br}(B\to X_s\gamma)=(3.23\pm0.41)\times 10^{-4}~,
\end{equation}
from the measurements of Belle \cite{Belle}, CLEO \cite{CLEO}, and ALEPH 
\cite{ALEPH} groups.
The $CP$ violating asymmetry in $B\to X_s\gamma$, defined by
\begin{equation}
A_{CP}(B\to X_s\gamma)=
 \frac{\Gamma({\bar B}\to X_s\gamma)-\Gamma(B\to X_{\bar s}\gamma)}
 {\Gamma({\bar B}\to X_s\gamma)+\Gamma(B\to X_{\bar s}\gamma)}~,
\end{equation}
is measured by CLEO \cite{CLEO2}:
\begin{equation}
A_{CP}(B\to X_s \gamma)=(-0.079\pm0.108\pm 0.022)(1.0\pm 0.030)~.
\end{equation}
The explicit expressions of the branching ratio and the $CP$ asymmetry are 
given in \cite{Kagan} in terms of the evolved Wilson coefficients at the
$\mu=m_b$ scale.
We adopt the constraints on $\xi$ established in \cite{KYL} from these 
experimental and theoretical results at $2\sigma$ C.L.:
\begin{eqnarray}
-0.002&<&{\rm Re}\xi_b+22|\xi_b|^2<0.0033~,\nonumber\\
-0.299&<&\frac{0.27{\rm Im}\xi_b}{0.095+12.54{\rm Re}\xi_b+414.23|\xi_b|^2}
<0.141~,\nonumber\\
|\xi_s|&<&0.012~.
\label{constraints}
\end{eqnarray}
Figure 1 shows the contour plot of $\lambda_\gamma$ for various values of
$\xi_{b,s}$.
Shaded ring denotes the allowed region by (\ref{constraints}).
Since the measured $CP$ asymmetry has rather large errors, constraints on 
$\xi$ are mainly from the ${\rm Br}(B\to X_s\gamma)$.
In Fig.\ 1, two contours for small and large value of $\xi_s$ are shown as an
illustration.
As one can expect from (\ref{lambdagamma}), deviation of $\lambda_\gamma$ from
$-1$ can be sizable if $\xi_s$ is large (Fig.\ 1 (a)).
On the other hand, if $\xi_s$ is very small, $\lambda_\gamma\approx -1$,
irrespective of $\xi_b$ (Fig.\ 1 (b)).
\par
As can be seen in Fig.\ 1 (a), large value of $\lambda_\gamma$ can be obtained
in the region of ${\rm Im}\xi_b=0$.
In Fig.\ 2, we give plots of $\lambda_\gamma$ vs $\xi_b$ for different values 
of $\xi_s$.
Here we assumed that $\xi_{b,s}$ are all real for simplicity.
Shaded bands are the allowed region by (\ref{constraints}).
\begin{figure}
\begin{center}
\epsfig{file=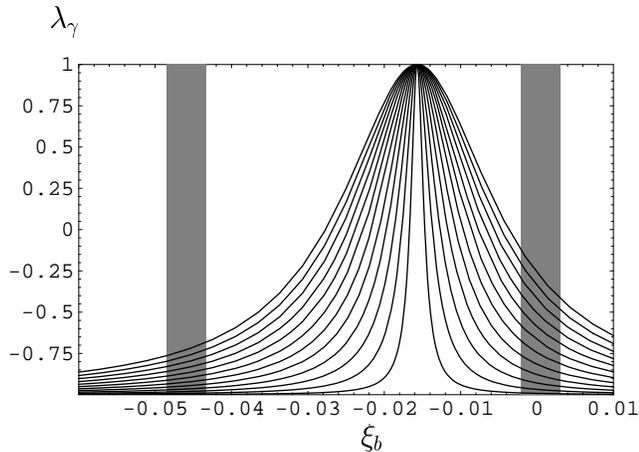,height=6cm}
\end{center}
\caption{Plots of $\lambda_\gamma$ as a function of $\xi_b$ for various $\xi_s$,
assuming that $\xi_{b,s}$ are real. 
Each curves corresponds to $\xi_s=0.001$, $0.002$, $\cdots$, $0.012$, from
bottom to top, respectively.
Shaded bands are the allowed region from $B\to X_s\gamma$.}
\end{figure}
In this case, a large deviation of $\lambda_\gamma$ from $-1$ is possible in 
the right-hand-side band.
This is quite natural since one can expect large $\lambda_\gamma$ in the region
of large $\xi_s$ and small $\xi_b$, by the inspection of (\ref{lambdagamma}).
We have
\begin{equation}
-1\le\lambda_\gamma\lesssim -0.12~.
\label{polbound}
\end{equation}
It should be noticed that current experimental bounds on $B\to X_s\gamma$ do 
not allow the different sign of $\lambda_\gamma$ compared to the SM prediction
at $2\sigma$ level.
Note that the upper bound of $\lambda_\gamma$ is chosen at the edge point 
of the allowed parameter space, say, $(\xi_b,\xi_s)=(-0.0021,0.012)$. 
If the new couplings are flavor-blind, i.e. $\xi_b=\xi_s$, then 
$\lambda_\gamma\simeq -0.96$ for $\xi_{b,s}=-0.002$.
Thus a large amount of reduction in $|\lambda_\gamma|$ implies that the 
anomalous right-handed couplings are flavor dependent.
\par
It is quite interesting to compare our result with that of the uMSSM 
\cite{Everett}.
The authors of Ref.\ \cite{Everett} proposed the "$C_{12}'$-dominated" scenario,
where the total contribution to $C_{12}$ is negligible while the main 
contribution to the ${\rm Br}(b\to s\gamma)$ is given by $C_{12}'$.
This is possible when the chargino, neutralino, and gluino contributions to
$C_{12}$ are canceled out by the $W$ and Higgs contributions.
Now that the size of $C_{12}$ is very small, they expect 
$\lambda_\gamma\approx +1$ as an extreme case, quite contrary to the SM 
predictions.
This is also very distinguishable from our result, since (\ref{polbound}) does
not allow the sign flip of $\lambda_\gamma$.
Thus the sign of $\lambda_\gamma$ is a very important landmark indicating
which kind of new physics is involved, if exists.
In case of ${\rm sgn}(\lambda_\gamma)>0$, models which produce only the 
anomalous right-hand top vertices would be disfavored.
At least we might need new particles, or new mechanism to make the $C_{12}'$
dominant.
\par
As introduced earlier, the integrated up-down asymmetry of 
$K_1(1400)(\to K^*\pi, \rho K)\to K^0\pi^+\pi^0$ or $K^+\pi^-\pi^0$ is reported 
to be $(0.33\pm0.05)\lambda_\gamma$ \cite{Gronau}, where
the uncertainty is a combined one from the uncertainties of $D$- and $S$-wave
amplitudes in the $K^*\pi$ channel, and $\rho K$ amplitude.
Within the SM where $\lambda_\gamma\approx -1$, it means that about 80
charged and neutral $B$ and ${\bar B}$ decays into $K\pi\pi\gamma$ are needed
to measure an asymmetry of $-0.33$ at $3\sigma$ level.
The authors of \cite{Gronau} estimated that at least $2\times 10^7~B{\bar B}$
pairs of both neutral and charged are required, with the use of 
${\rm Br}(B\to K_1(1400)\gamma)=0.7\times 10^{-5}$ and 
${\rm Br}(K_1(1400)\to K^*\pi)=0.94\pm 0.06$.
\par
If the anomalous right-handed couplings were present, we would need more 
$B{\bar B}$ pairs because new couplings will reduce the value of 
$|\lambda_\gamma|$.
For example, in case of $\lambda_\gamma=-0.5$, we need 4 times larger number
of $B{\bar B}$ pairs ($8\times 10^7$) to see the $3\sigma$ deviation of the
up-down asymmetry from both zero and the SM prediction.
Fortunately, this is already within the reach of current $B$ factories 
\cite{BABAR}.
If the deficiency of the up-down asymmetry were found, then the direct searches
of the anomalous top couplings at the LHC in coming years would be very 
exciting to check the consistency.
\par
In conclusion, we have analyzed the effects of anomalous right-handed top
couplings on the photon polarization in radiative $B\to K_{\rm res}\gamma$ 
decays.
The photon polarization parameter $\lambda_\gamma$ defined by the ratio of
the relevant Wilson coefficients is a useful observable for measuring the photon
helicity.
It is found that the new couplings can reduce $|\lambda_\gamma|$ significantly,
compared to the SM prediction of $\lambda_\gamma\approx -1$,
while satisfying the strong constraints from the measured branching ratio and 
$CP$ asymmetry of $B\to X_s\gamma$.
We also find that the anomalous right-handed top couplings would not produce 
different sign of $\lambda_\gamma$ from the SM prediction.
This is a crucial point to distinguish our case from other scenarios such as 
uMSSM where an extreme value of $\lambda_\gamma=+1$ can be possible through
the "$C_{12}'$-dominated" mechanism.
The importance of present work also lies in the fact that current $B$ factories 
can produce enough $B{\bar B}$ pairs to analyze the photon polarization in
$B\to K_{\rm res}\gamma$.
\par
The author is grateful for Kang-Young Lee's valuable comments on the Wilson
coefficients. 
He also gives thanks to Heyoung Yang for helpful discussions.
This work was supported by the BK21 program of the Korean Ministry of Education.




\begin{thebibliography}{99}
\bibitem{Song}
I.S.\ Choi, S.Y.\ Choi, and H.S.\ Song, \prd{41}{1990}{1695}.
\bibitem{Atwood}
D.\ Atwood, M.\ Gronau, and A.\ Soni, \prl{79}{1997}{185}.
\bibitem{Mannel}
T.\ Mannel and S.\ Recksiegel, Acta Phys. Pol. B {\bf 28}, 2489 (1997);
G.\ Hiller and A.\ Kagan, \prd{65}{2002}{074038}.
\bibitem{Melikhov}
D.\ Melikhov, N.\ Kikitin, and S.\ Simula, \plb{442}{1998}{381};
F.\ Kr\"uger, L.M.\ Sehgal, N.\ Sinha, and R.\ Sinha, \prd{61}{2000}{114028};
C.S.\ Kim, Y.G.\ Kim, C.D.\ L\"u, and T.\ Morozumi, \prd{62}{2000}{034013};
Y.\ Grossman and D.\ Pirjol, \jhep{06}{2000}{029}.
\bibitem{Gronau}
M.\ Gronau, Y.\ Grossman, D.\ Pirjol, and A.\ Ryd, \prl{88}{2002}{051802};
M.\ Gronau and D.\ Pirjol, \prd{66}{2002}{054008}.
\bibitem{Babu}
K.S.\ Babu. K.\ Fujikawa, and A.\ Yamada, \plb{333}{1994}{196};
P.\ Cho and M.\ Misiak, \prd{49}{1994}{5894}.
\bibitem{Everett}
L.\ Everett, G.L.\ Kane, S.\ Rigolin, L.T.\ Wang, T.T.\ Wang, 
\jhep{01}{2002}{022}.
\bibitem{KYL}
J.-P.\ Lee and K.Y.\ Lee, \epjc{29}{2003}{373}.
\bibitem{Inami}
T.\ Inami and C.S.\ Lim, Prog. Theor. Phys. {\bf 65}, 297 (1981);
G.\ Buchalla, A.J.\ Buras, M.E.\ Lautenbacher, \rmp{68}{1996}{1125};
A.J.\ Buras, hep-ph/9006471.
\bibitem{Cho}
P.\ Cho and M.\ Misiak, \prd{49}{1994}{5894};
K.Y.\ Lee and W.Y.\ Song, \prd{66}{2002}{057901}.
\bibitem{Belle}
H.\ Tajima \etal, Belle Collaboration, \ijmpa{17}{2002}{2967}.
\bibitem{CLEO}
S.\ Chen \etal, CLEO Collaboration, \prl{87}{2001}{251807}.
\bibitem{ALEPH}
R.\ Barate \etal, ALEPH Collaboration, \plb{429}{1998}{169}.
\bibitem{CLEO2}
S.\ Chen \etal, CLEO Collaboration, \prl{86}{2001}{5661}.
\bibitem{Kagan}
A.\ Kagan and M.\ Neubert, \epjc{7}{1999}{5}; \prd{58}{1998}{094012}.
\bibitem{BABAR}
See, for example, B.\ Aubert \etal, BABAR Collaboration, hep-ex/0308021.

\end{thebibliography}
\end{document}